\documentclass[12pt]{article}
\usepackage{graphicx}
\usepackage{natbib} 
\usepackage{url} 

\usepackage{amsmath}
\usepackage{amssymb}
\usepackage{amsfonts}
\usepackage{multirow}
\usepackage{amsthm}
\usepackage{algorithm}
\usepackage{algpseudocode}

\usepackage{moreverb,url}

\usepackage{graphicx}
\usepackage{amsmath}
\usepackage{times}
\usepackage{bm}
\usepackage{enumerate}
\usepackage{mathtools}
\usepackage{setspace}
\usepackage{array,booktabs}
\usepackage{tikz}
\usepackage{natbib}
\usepackage{mathtools}
\usepackage{chngcntr}
\usepackage{algorithm}
\usepackage[utf8]{inputenc}
\usepackage{xcolor, soul}
\usepackage{hyperref}

\usepackage{graphicx}
\usepackage{amsmath}

\usepackage{times}
\usepackage{bm}
\usepackage{enumerate}
\usepackage{xurl}

\usepackage{mathtools}
\usepackage{setspace}
\usepackage{array,booktabs}
\usepackage{tikz}
\usepackage{mathtools}
\usepackage{chngcntr}

\theoremstyle{definition}

\newcommand{\blind}{1}

\usepackage[margin=0.7in]{geometry}

\begin{document}

\def\spacingset#1{\renewcommand{\baselinestretch}%
{#1}\small\normalsize} \spacingset{1}

\date{}

\if1\blind
{
	\title{\bf A Class of Computational Methods to Reduce Selection Bias when Designing Phase 3 Clinical Trials}
	\author{
		Tianyu Zhan\thanks{Tianyu Zhan is an employee of AbbVie Inc. Corresponding author email address: \texttt{tianyu.zhan.stats@gmail.com}.} \\
		\footnotesize Data and Statistical Sciences, AbbVie Inc., North Chicago, IL, USA}
	\maketitle
} \fi

\if0\blind
{
	\title{\bf A Class of Computational Methods to Reduce Selection Bias when Designing Phase 3 Clinical Trials}
	\maketitle
} \fi

\bigskip
\begin{abstract}
When designing confirmatory Phase 3 studies, one usually evaluates one or more efficacious and safe treatment option(s) based on data from previous studies. However, several retrospective research articles reported the phenomenon of ``diminished treatment effect in Phase 3'' based on many case studies. Even under basic assumptions, it was shown that the commonly used estimator could substantially overestimate the efficacy of selected group(s). As alternatives, we propose a class of computational methods to reduce estimation bias and mean squared error (MSE) with a broader scope of multiple treatment groups and flexibility to accommodate summary results by group as input. Based on simulation studies and a real data example, we provide practical implementation guidance for this class of methods under different scenarios. For more complicated problems, our framework can serve as a starting point with additional layers built in. Proposed methods can also be widely applied to other selection problems.
\end{abstract}

\noindent%
\vfill
\noindent%

\newpage
\spacingset{2} 

\section{Introduction}

In clinical drug development, confirmatory Phase 3 studies are usually conducted to comprehensively evaluate the safety and efficacy of the study drug after exploratory Phase 2 studies \citep{e8}. To properly design Phase 3 studies, one needs to accurately characterize the efficacy profile of one or more selected efficacious and safe treatment option(s) to inform many key decisions, for example, Go/No-Go and sample size calculation. However, quite a few retrospective research articles reported the phenomenon of ``diminished treatment effect in Phase 3'': FDA studied 22 recent cases in which promising Phase 2 results were not confirmed in Phase 3 and found 21 of them were due to lack of efficacy \citep{fda2017}; treatment effect sizes of progression-free survival (PFS) were on average $26\%$ larger in Phase 2 as compared with Phase 3 in 57 pairs of oncology studies \citep{liang2019comparison}; 35 out of 43 Phase 3 studies of chemotherapy in advanced solid malignancies had lower response rates than preceding Phase 2 studies \citep{zia2005comparison}. 

This question concerning the efficacy gap between previous studies and Phase 3 studies is important, but is also challenging to resolve. There are several caveats that may contribute to this gap, for example, temporal drift due to the standard of care improvements or other factors \citep{saville2022bayesian}, patient heterogeneity across studies \citep{liang2019comparison}, variability in results based on limited sample size. As a starting point, we consider a typical approach of directly choosing the treatment group(s) with the best outcome(s) based on previous studies, and use the corresponding results as assumptions for Phase 3 design. Under a basic scenario where true response means between studies are the same for a selected group, this estimator may substantially overestimate its response mean in Phase 3 with insufficient power, as further discussed in later sections, including a toy simulation in Table \ref{T:toy}. As discussed in Section \ref{sec:dis}, this framework can be extended with additional layers to handle more complicated problems, e.g., temporal drift, and other base estimators, e.g., the minimum efficacious dose (MED) modeled from MCP-Mod \citep{bretz2005combining}.

There were some previous theoretical works conducted to study this problem of estimating the larger of two means for some specific distributions. \cite{blumenthal1968estimation} and \cite{dahiya1974estimation} investigated this under two Normal distributions and a common known variance, and showed that no unbiased estimator could exist \citep{blumenthal1968estimation}. This phenomenon of the non-existence of unbiased estimators was further studied in more generalized distributions from two groups, for example, Normal distributions with common but unknown variance \cite{hsieh1981estimating}, a general class of distributions \citep{ishwaei1985non}, double exponential distributions with unknown locations \citep{kumar1993unbiased}. On the other hand, unbiased estimators may exist under some special settings, e.g., two gamma distributions with a common and known shape parameter \citep{vellaisamy1988estimation}. As a general approach, \cite{rosenkranz2014bootstrap} proposed to correct bias using non-parametric Bootstrap \citep{efron1994introduction, davison1997bootstrap, kosmidis2014bias} to accommodate general distribution assumptions from two groups. However, patient-level data are needed to implement this method. In this article, we consider a more general scope of "previous studies", in the sense that it can be in-house Phase 2 studies with patient-level data under multiple doses and/or multiple compounds, or external studies with only summary data available to characterize assumptions of the active comparator(s) in the new Phase 3 study. Additionally, it is also common to have more than two treatment options to be selected in the design of Phase 3 trials. 

{Additionally, there are several methods proposed to correct selection bias in clinical trials with two or more stages. Based on} \cite{whitehead1986bias}, \cite{stallard2005point} {developed an iterative approach to reduce the estimation bias conditional on the selection of a treatment group. This method requires analytic derivation of the conditional bias given a specific setting, e.g., equal-variance considered in} \cite{stallard2005point}. {The single and double Bootstrap methods introduced in Section} \ref{sec:method_boot} {have a similar idea of correcting bias iteratively, but utilizes empirical Bootstrap distributions to estimate the bias. Our proposed approaches are also more general to cover settings with unequal-variance. \cite{bauer2010selection} {investigated the bias and MSE when estimating the efficacy of the best treatment in multi-stage trials with sample size adaptation and homogeneous variance. They emphasized that the quantification of the bias is possible only in designs with planned adaptivity} \citep{bauer2010selection}. \cite{hwang1993empirical} {and} \cite{Lindley1962} {proposed a shrinkage estimator with superior performance of Bayes risk as compared with the typical maximum-likelihood estimator (MLE). This shrinkage estimator is briefly reviewed in Section} \ref{sec:shrink} {with comparison results in Section} \ref{sec:sim}. Two recent papers nicely reviewed point estimation for adaptive trial designs, including bias reduction in multi-arm multi-stage designs with treatment selection \citep{robertson2023point1, robertson2023point2}.

{The motivation of our proposed methods is to empirically estimate the bias for correction with computational approaches, such as Bootstrap} \citep{efron1994introduction} {or Jackknife} \citep{quenouille1949approximate}. {This framework can naturally accommodate general settings, such as multiple (more than two) groups based on either subject-level data or group-level summary data, homogeneous or heterogeneous variance between treatment groups. Our scope is broader to cover typical Phase 2 studies with patient-level data, and external studies with only summary data available based on literature. As compared with single Bootstrap methods, double Bootstrap methods can further reduce bias with slightly larger mean squared error (MSE) and an additional cost of computation, with results in Section} \ref{sec:sim}. {We further propose hybrid estimators based on double Bootstrap estimators and shrinkage estimators} \citep{hwang1993empirical, Lindley1962} {to balance the reductions in both bias and MSE.}

The remainder of this article is organized as follows. In Section \ref{sec:setup}, we introduce the setup of this problem and notations. In Section \ref{sec:method}, a class of computational methods is proposed, and an existing shrinkage estimator \citep{hwang1993empirical, Lindley1962} is reviewed. Simulation studies are performed in Section \ref{sec:sim} to demonstrate the potential gains of those proposed methods in terms of bias and MSE under different settings. We apply our methods to a Phase 2/3 seamless trial in Section \ref{sec:real}. Discussions are provided in Section \ref{sec:dis}. 

\section{Setup}
\label{sec:setup}

Consider a previous study with $I$ active treatment groups and $n_i$ patients randomized to the $i$th treatment group, for $i = 1, \cdots, I$. We consider the response $X_{i, j}$ of the treatment group $i$, for $i = 1, \cdots, I$, and the subject $j$, for $j = 1, \cdots, n_i$, follows a Normal distribution,
\begin{equation}
	\label{equ:normal}
	X_{i, j} \sim \mathcal{N}\left(\theta_i, \sigma_i^2\right),
\end{equation}
where $\theta_i$ is the mean and $\sigma_i$ is the standard deviation of the treatment group $i$. We assume that a larger value of $X_{ij}$ corresponds to a better outcome, {and $\sigma_i$ is unknown}. Denote $\boldsymbol{X}_i = \left(X_{i, 1}, \cdots, X_{i, n_i} \right)$ as the vector of responses from group $i$, and $\boldsymbol{X} = (\boldsymbol{X}_1, \cdots, \boldsymbol{X}_I)$. 

{After obtaining data from multiple treatment groups, the study team will usually select one or two treatment group(s) to confirm findings in Phase 3 studies.} We consider a motivating scenario where all treatment groups have similar safety profiles, and the most efficacious group will be moved to Phase 3. A key question is how to accurately characterize the efficacy of this selected group for sample size calculation. 

The corresponding statistical question is to use observed data $\left( \boldsymbol{X}_1, \cdots, \boldsymbol{X}_I \right)$ to estimate the parameter of interest $\theta_{max}$, defined as,
\begin{equation}
	\label{equ:theta_max}
	\theta_{max} = \max(\theta_1, \cdots, \theta_I). 
\end{equation}
A traditional estimator $\widehat{\theta}$ is commonly used in practice to estimate $\theta_{max}$:
\begin{equation}
	\label{equ:theta_naive}
	\widehat{\theta}(\boldsymbol{X}) = \max\left[\widetilde{\theta}(\boldsymbol{X}_1), \cdots, \widetilde{\theta}(\boldsymbol{X}_I)\right], 
\end{equation}
where $\widetilde{\theta}(x)$ is the sample mean of $x$, and $\widetilde{\theta}(\boldsymbol{X}_i)$ as an unbiased estimator of $\theta_i$. However, $\widehat{\theta}(\boldsymbol{X})$ may overestimate $\theta_{max}$ in finite-samples. Even though $\widetilde{\theta}(\boldsymbol{X}_i)$ can accurately estimate $\theta_i$ with no bias for each treatment group $i$, one does not know which treatment group has the highest true response mean $\theta_i$ in (\ref{equ:theta_max}) based on observed data. 

To provide a numerical illustration of bias, we conduct a toy simulation with $I=2$ treatment groups, $\theta_1=0.9$, $\theta_2 = 1$, and $\sigma_1 = \sigma_2 = 5$ under three magnitudes of sample size $n$. Table \ref{T:toy} shows that the traditional estimator $\widehat{\theta}$ can overestimate $\theta_{max}$ by $40\%$ under a moderate sample size $n=40$, but the bias shrinks as $n$ increases. When $n = 40,000$, the probability of selecting the correct treatment group $i=2$ is nearly $100\%$, and therefore, $\widehat{\theta}(\boldsymbol{X})$ is close to $\widetilde{\theta}(\boldsymbol{X}_2)$ as an unbiased estimator of $\theta_{max} = \theta_2$. 

\begin{table}[ht]
	\centering
	\begin{tabular}{ccccc}
		$\theta_{max}$ & $n$ & $E\big(\widehat{\theta} \big)$ & $E(\widehat{\theta} ) - \theta_{max}$ & Prob of correctly selecting $i=2$ \\ 
		\hline
		1 & 40 & 1.40 & 0.40 & 0.53 \\ 
		& 4000 & 1.01 & 0.01 & 0.82 \\ 
		& 40000 & 1.00 & 0.00 & 1.00 \\ 
		\hline
	\end{tabular}
	\caption{A toy simulation to evaluate the bias of $\widehat{\theta}$ when estimating $\theta_{max}$. }
	\label{T:toy}
\end{table}

\section{Proposed Methods}
\label{sec:method}

In this section, we introduce a class of computational methods based on Bootstrap or Jackknife principles to reduce estimation bias. 

\subsection{Single and Double Bootstrap}
\label{sec:method_boot}

Suppose that we have $\widehat{\theta}(\boldsymbol{X})$ in (\ref{equ:theta_naive}) as an initial estimator of $\theta_{max}$. Its bias at $\theta_0$ is denoted as $A(\theta_0)$,
\begin{equation}
	\label{equ:bias_naive}
	A(\theta_0) = E\left[\widehat{\theta}(\boldsymbol{X}) \right] - \theta_0.
\end{equation}
Since the true value $\theta_0$ of $\theta_{max}$ is to be estimated and the functional form of $A(\cdot)$ is usually unknown, one can use $\widehat{A}\left[\widehat{\theta}(\boldsymbol{X}) \right]$ to approximate $A(\theta_0)$,
\begin{equation}
	\label{equ:bias_naive_est}
	\widehat{A}\left[\widehat{\theta}(\boldsymbol{X}) \right] = \widehat{E}\left[\widehat{\theta}(\boldsymbol{X}_B) \right] - \widehat{\theta}(\boldsymbol{X}),
\end{equation}
where $\widehat{E}$ is the empirical expectation based on Monte Carlo Bootstrap data $\boldsymbol{X}_B$ with size $B$. The single Bootstrap estimator $\widehat{\theta}^{(1)}(\boldsymbol{X})$ \citep{efron1994introduction, davison1997bootstrap, kosmidis2014bias} can then be constructed as,
\begin{equation}
	\label{equ:Boot_single}
	\widehat{\theta}^{(1)}(\boldsymbol{X}) = \widehat{\theta}(\boldsymbol{X}) - \widehat{A}\left[\widehat{\theta}(\boldsymbol{X}) \right]. 
\end{equation}

Figure \ref{F:Boot} left-hand side provides a graphical illustration of the construction above. Algorithm \ref{alg:boot_single} streamlines the workflow to compute $\widehat{\theta}^{(1)}(\boldsymbol{X})$ based on $B$ Bootstrap samples. 

To further reduce bias, we can iteratively apply the above approach with $\widehat{\theta}^{(1)}(\boldsymbol{X})$ as the initial estimator to obtain the double Bootstrap estimator $\widehat{\theta}^{(2)}(\boldsymbol{X})$ as in Figure \ref{F:Boot} right-hand side. Algorithm \ref{alg:boot_double} demonstrates that the computation of $\widehat{\theta}^{(2)}(\boldsymbol{X})$ requires $B^2$ Bootstrap samples. This strategy is analog to the calibration of Bootstrap to obtain second-order accurate confidence intervals \citep{efron1994introduction}. Based on our simulation studies in Section \ref{sec:sim}, $\widehat{\theta}^{(2)}(\boldsymbol{X})$ has a satisfactory finite-sample performance in terms of bias and MSE. The triple (or even a higher-order) Bootstrap estimator can also be implemented to seek potential improvements, but with a cost of a much heavier computational burden. Section \ref{sec:sen_boot} provides more discussion on higher-order Bootstrap estimators. 

Next we provide more details on simulating Bootstrap samples from data. Taking the single Bootstrap as an example, our strategy is to resample $\boldsymbol{X}_b^\ast$ from observed data $\boldsymbol{X}$ blocked by groups. To be more specific, for each treatment group $i$, we generate Bootstrap samples $\boldsymbol{Y}_{b, i}$ of size $n_i$ from $\boldsymbol{X}_i$, and then obtain $\boldsymbol{X}_b^\ast = \left(\boldsymbol{Y}_{b, 1}, \cdots, \boldsymbol{Y}_{b, I} \right)$. For sampling methods, one can adopt the Nonparametric Bootstrap (NB) to sample $n_i$ observations from $\boldsymbol{X}_i$ with replacement to get $\boldsymbol{Y}_{b, i}$, as considered in \cite{rosenkranz2014bootstrap}. An alternative approach is the Parametric Bootstrap (PB) with distribution assumptions, for example, a Normal distribution with sample mean $\widetilde{\theta}(\boldsymbol{X}_i)$ as the mean parameter and {empirical standard deviation $\widetilde{\sigma}(\boldsymbol{X}_i)$ as the standard deviation parameter}. PB is flexible to cover scenarios where only summary statistics (e.g., $n$, $\widetilde{\theta}$, $\widetilde{\sigma})$ for each group are reported in the literature or other external sources. In Section \ref{sec:sim}, we have scenarios with mixture distributions to evaluate the robustness of PB. 

Now we have four Bootstrap estimators: $\widehat{\theta}^{(1)}_{PB}$, $\widehat{\theta}^{(2)}_{PB}$, $\widehat{\theta}^{(1)}_{NB}$, $\widehat{\theta}^{(2)}_{NB}$, where the superscript ``$(1)$'' corresponds to the single Bootstrap, ``$(2)$'' to the double Bootstrap, the subscript ``$PB$'' to Parametric Bootstrap, and ``$NB$'' to Nonparametric Bootstrap. 

\begin{figure}[h]
	\centering
	\includegraphics[width=16cm]{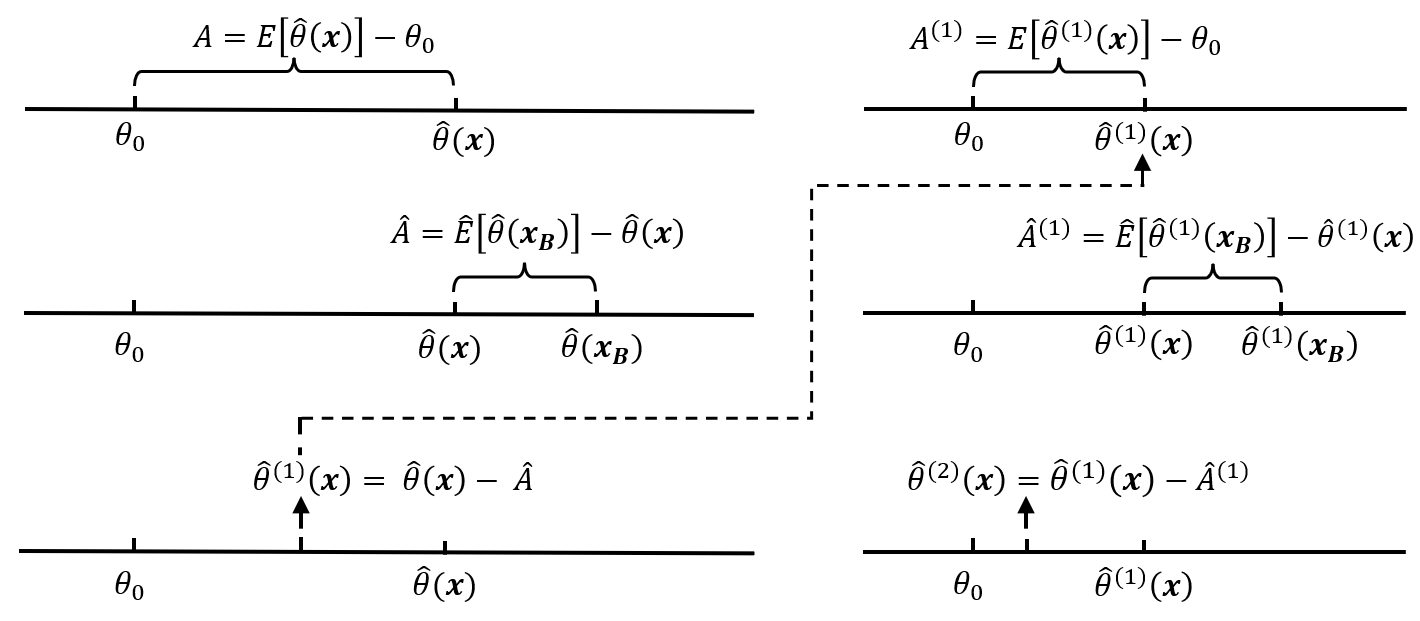}
	\caption{Graphical illustration of the single Bootstrap estimator $\widehat{\theta}^{(1)}$ (left) and the double Bootstrap estimator $\widehat{\theta}^{(2)}$ (right). }
	\label{F:Boot}
\end{figure}

\begin{algorithm}
	\caption{Single Bootstrap}
	\label{alg:boot_single}
	\begin{algorithmic}
		\State \textbf{Input: } $\boldsymbol{X}$
		\State
		\State \textbf{Procedure:}
		\State \quad \textbf{For} Bootstrap index $b$ from $1$ to $B$, \textbf{Do}: 
		\State \quad\quad \textbf{Simulate} Bootstrap sample $\boldsymbol{X}_b^\ast$ based on $\boldsymbol{X}$ 
		\State \quad \textbf{End}
		\State \quad \textbf{Compute} $\widehat{A}\left[\widehat{\theta}(\boldsymbol{X}) \right] = \sum_{b=1}^B \widehat{\theta}(\boldsymbol{X}_b^\ast) / B - \widehat{\theta}(\boldsymbol{X})$
		\State
		\State \textbf{Output: } $\widehat{\theta}^{(1)}(\boldsymbol{X}) = \widehat{\theta}(\boldsymbol{X}) - \widehat{A}\left[\widehat{\theta}(\boldsymbol{X}) \right]$
	\end{algorithmic}
\end{algorithm}

\begin{algorithm}
	\caption{Double Bootstrap}
	\label{alg:boot_double}
	\begin{algorithmic}
		\State \textbf{Input: } $\boldsymbol{X}$
		\State
		\State \textbf{Procedure:}
		\State \quad \textbf{For} Bootstrap index $b$ from $1$ to $B$, \textbf{Do}:
		\State \quad\quad \textbf{Simulate} Bootstrap sample $\boldsymbol{X}_b^\ast$ based on $\boldsymbol{X}$
		\State \quad\quad\quad \textbf{For} Bootstrap index $c$ from $1$ to $B$, \textbf{Do}:
		\State \quad\quad\quad\quad \textbf{Simulate} Bootstrap sample $\boldsymbol{X}_{b, c}^{\ast\ast}$ based on $\boldsymbol{X}_b^\ast$
		\State \quad\quad\quad \textbf{End}
		\State \quad\quad \textbf{Compute} $\widehat{A}\left[\boldsymbol{X}_b^\ast\right] = \sum_{c=1}^B \widehat{\theta}(\boldsymbol{X}_{b, c}^{\ast\ast}) / B - \widehat{\theta}(\boldsymbol{X}_b^\ast)$
		\State \quad\quad \textbf{Obtain} $\widehat{\theta}^{(1)}(\boldsymbol{X}_b^\ast) = \widehat{\theta}(\boldsymbol{X}_b^\ast) - \widehat{A}\left[\boldsymbol{X}_b^\ast\right]$
		\State \quad \textbf{End}
		\State \quad \textbf{Compute} $\widehat{A}\left[\widehat{\theta}^{(1)}(\boldsymbol{X}) \right] = \sum_{b=1}^B \widehat{\theta}^{(1)}(\boldsymbol{X}_b^\ast) / B - \widehat{\theta}^{(1)}(\boldsymbol{X})$
		\State
		\State \textbf{Output: } $\widehat{\theta}^{(2)}(\boldsymbol{X}) = \widehat{\theta}^{(1)}(\boldsymbol{X}) - \widehat{A}\left[\widehat{\theta}^{(1)}(\boldsymbol{X}) \right]$
	\end{algorithmic}
\end{algorithm}

\subsection{Jackknife}

The Jackknife is a well-established technique to correct bias \citep{quenouille1949approximate, miller1974jackknife}. Suppose that the bias of $\widehat{\theta}(\boldsymbol{X})$ under sample size $n$ can be expressed as, 
\begin{equation}
	\label{equ:bias_JK_1}
	E\left[\widehat{\theta}(\boldsymbol{X}); n \right] - \theta_0 = \frac{a_1}{n} + \frac{a_2}{n^2} + \mathcal{O}(n^{-3}),
\end{equation}
where $a_1$ and $a_2$ are unknown and do not depend on $n$. The bias with a sample size of $n-1$ is,
\begin{equation}
	\label{equ:bias_JK_2}
	E\left[\widehat{\theta}(\boldsymbol{X}); n-1 \right] - \theta_0 = \frac{a_1}{n-1} + \frac{a_2}{(n-1)^2} + \mathcal{O}(n^{-3}).
\end{equation}

In order to correct the order $1/n$ term, one can construct the following Jackknife estimator $\widehat{\theta}_{JK}(\boldsymbol{X})$,
\begin{equation}
	\label{equ:JK}
	\widehat{\theta}_{JK}(\boldsymbol{X}) = n \widehat{\theta}(\boldsymbol{X}) - (n-1) \widehat{\theta}_{(\bullet)}(\boldsymbol{X}),
\end{equation}
where $\widehat{\theta}_{(\bullet)}(\boldsymbol{X}) = {\sum_{j=1}^n\widehat{\theta}\left(\boldsymbol{X}_{-j}\right)}/{n}$, and $\boldsymbol{X}_{-j}$ is $\boldsymbol{X}$ with $j$th observation deleted. 

It can be shown that the bias of $\widehat{\theta}_{JK}(\boldsymbol{X})$ is now with an order of $1/n^2$ \citep{miller1974jackknife}. A graphical illustration of this bias reduction is provided in Figure \ref{F:Jack}. As compared with Bootstrap methods, $\widehat{\theta}_{JK}(\boldsymbol{X})$ is computationally friendly, and can also give exact results without a need to specify random seeds. 

\begin{figure}[h]
	\centering
	\includegraphics[width=14cm]{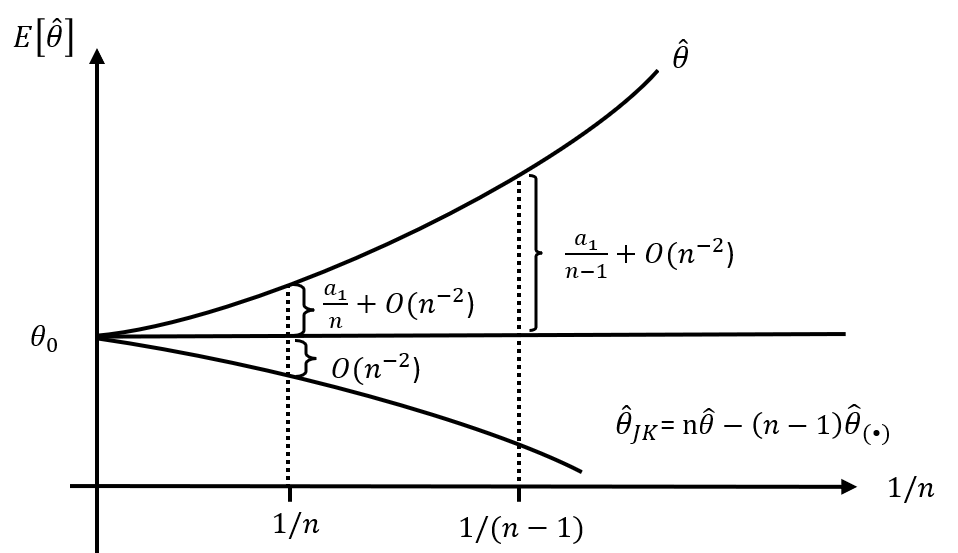}
	\caption{Graphical illustration of the Jackknife estimator $\widehat{\theta}_{JK}$.}
	\label{F:Jack}
\end{figure}

\subsection{Shrinkage Estimators}
\label{sec:shrink}

\cite{hwang1993empirical} {and} \cite{Lindley1962} {considered the following shrinkage estimator $\widehat{\theta}_{S}(\boldsymbol{X})$ for $\theta_{max}$ to reduce MSE.} 
\begin{align}
	\widehat{\theta}_{S}(\boldsymbol{X}) & = C_{+} \widehat{\theta}(\boldsymbol{X}) + \left(1-C_{+}\right) \widetilde{\theta}(\boldsymbol{X}) \label{equ:shrinkage} \\
	C_{+} & = \max(0, C) \nonumber\\
	C & = 1- \frac{(I-1)\sigma^2}{\sum_{i=1}^I n_i\left[\widetilde{\theta}(\boldsymbol{X}_i) - \widetilde{\theta}(\boldsymbol{X})\right]^2} \label{equ:shrink_c}
\end{align}
{Their original estimator is based on a setting of common and known variance $\sigma^2$ for each treatment group $i$ in} (\ref{equ:normal}). {For evaluation in this article, we replace $\sigma^2$ in} (\ref{equ:shrink_c}) {by the average of empirical variance estimators from all $I$ groups. Moreover, the constant $(I-1)$ is modified from $(I-3)$ for $I\geq 4$ in} \cite{hwang1993empirical} {and} \cite{Lindley1962} {to accommodate a general setting with $I \geq 2$ as suggested by} \cite{carreras2013shrinkage}. {Intuitively, when $\theta_1, \cdots, \theta_I$ are far from each other, $C$ in} (\ref{equ:shrink_c}) {will be close to $1$, because $\sum_{i=1}^I n_i\left[\widetilde{\theta}(\boldsymbol{X}_i) - \widetilde{\theta}(\boldsymbol{X})\right]^2$ is relatively large. The estimator $\widehat{\theta}_{S}(\boldsymbol{X})$ will be close to $\widehat{\theta}(\boldsymbol{X})$ with small bias under this setting} \citep{carreras2013shrinkage}. {Otherwise, when $\theta_1, \cdots, \theta_I$ are close to each other, $\widehat{\theta}_{S}(\boldsymbol{X})$ will be close to $\widetilde{\theta}(\boldsymbol{X})$ as the overall mean. Superior performance of $\widehat{\theta}_{S}(\boldsymbol{X})$ in terms of Bayes risk was studied in} \cite{hwang1993empirical} {and} \cite{carreras2013shrinkage}.

\subsection{Hybrid Estimators based on Double Bootstrap and Shrinkage}

{The double bootstrap estimators $\widehat{\theta}^{(2)}_{PB}$ and $\widehat{\theta}^{(2)}_{NB}$ may further reduce estimation bias as compared with their single Bootstrap versions, but with a potential cost of increased MSE. This phenomenon was observed in some previous works} \citep{hsu1986monte, mackinnon1998approximate, ouysse2011computationally}, {and our simulation results in Section} \ref{sec:sim}.  

{In this article, we also consider a natural generalization to ensemble double Bootstrap estimators and shrinkage estimators in Section} \ref{sec:shrink}, {with a goal to balance reductions in bias and MSE. The proposed hybrid estimators $\widehat{\theta}^{(2)}_{PB, S}$ and $\widehat{\theta}^{(2)}_{NB, S}$ substitute $\widehat{\theta}$ in} (\ref{equ:shrinkage}) {by $\widehat{\theta}^{(2)}_{PB}$ and $\widehat{\theta}^{(2)}_{NB}$, respectively. }

\section{Simulation}
\label{sec:sim}

\subsection{Main Study}
{In this section, we conduct simulations with $I=3$ treatment groups and $n=40$ per group to evaluate the performance of several existing estimators $\widehat{\theta}$,} $\widehat{\theta}_S$ \citep{hwang1993empirical, Lindley1962, carreras2013shrinkage} and $\widehat{\theta}^{(1)}_{NB}$ \citep{rosenkranz2014bootstrap} {and our proposed estimators $\widehat{\theta}^{(2)}_{NB}$, $\widehat{\theta}^{(2)}_{NB, S}$, $\widehat{\theta}^{(1)}_{PB}$, $\widehat{\theta}^{(2)}_{PB}$,
	$\widehat{\theta}^{(2)}_{PB, S}$ and
	$\widehat{\theta}_{JK}$. An additional setting of $I=4$ is considered in Table} \ref{sim_tab_uncond_four}. {Denote $\boldsymbol{\theta} = (\theta_1, \theta_2, \theta_3)$ as the response mean vector, and $\boldsymbol{\sigma} = (\sigma_1, \sigma_2, \sigma_3)$ as the standard deviation vector. The number of Bootstrap samples is set at $B=80$, and the number of simulation iterations is $10,000$. Discussion on how to choose the value of $B$ is provided in Section} \ref{sec:sen_boot}. 

We consider the following four simulation scenarios.
\begin{itemize}
	\item S1: Varying mean vector $\boldsymbol{\theta}$ at $(1, 1, 1)$, $(1, 1, 1.2)$, $(1, 1.1, 1.2)$ and $(1, 1.2, 1.2)$ with $\boldsymbol{\sigma} = (5, 5, 5)$ and Normal distribution of $\boldsymbol{X}$
	\item S2: Varying mean vector $\boldsymbol{\theta}$ as in S1 but with $\boldsymbol{\sigma} = (3, 4, 5)$ and Normal distribution of $\boldsymbol{X}$
	\item S3: Varying $w$ at $0.1$, $0.2$, $0.3$ and $0.5$ with $\boldsymbol{\sigma} = (5, 5, 5)$, $\boldsymbol{\theta} = (1, 1.1, 1.2)$ and a mixture of Gamma and Normal distributions of $\boldsymbol{X}$
	\item S4: Varying $w$ at $0.1$, $0.2$, $0.3$ and $0.5$ with $\boldsymbol{\sigma} = (5, 5, 5)$, $\boldsymbol{\theta} = (1, 1.1, 1.2)$ and a mixture of Uniform and Normal distributions of $\boldsymbol{X}$
\end{itemize}
S1 considers different values of $\boldsymbol{\theta}$ under homogeneous $\boldsymbol{\sigma} = (5, 5, 5)$, while S2 is for heterogeneous $\boldsymbol{\sigma} = (3, 4, 5)$. S3 evaluates estimators based on a mixture distribution with $w$ ($w \in [0,1]$) proportion of Gamma distribution and $1-w$ proportion of Normal distribution for each treatment group. S4 studies Uniform distribution as the outlier distribution. The shape and scale parameters of Gamma distributions or the minimum and maximum parameters of Uniform distributions are specified to match the mean and standard deviation for each treatment group. Those  Parametric Bootstrap estimators (i.e.,  $\widehat{\theta}^{(1)}_{PB}$, $\widehat{\theta}^{(2)}_{PB}$, $\widehat{\theta}^{(2)}_{PB, S}$) still use Normal distribution as the re-sampling assumption. S3 and S4 essentially evaluate the robustness of different estimators based on data sampling distributions deviating from the Normal assumption.

{Table} \ref{sim_tab_uncond} {evaluates the unconditional or marginal bias and MSE of those estimators when estimating $\boldsymbol{\theta}_{max}$. Among existing estimators, $\widehat{\theta}_{S}$ and $\widehat{\theta}^{(1)}_{NB}$ can generally reduce bias and MSE as compared with the traditional estimator $\widehat{\theta}$. For our proposed estimators, both $\widehat{\theta}^{(2)}_{PB}$ and $\widehat{\theta}^{(2)}_{NB}$ can substantially reduce estimation bias but with increased MSE. As a better balance between bias reduction and MSE reduction, our hybrid parameters $\widehat{\theta}^{(2)}_{PB, S}$ and $\widehat{\theta}^{(2)}_{NB, S}$ have the smallest MSE, and also smaller bias than three existing estimators. The Normal assumption for PB is usually reasonable to assume for problems with response mean as the parameter of interest and a moderate sample size} \citep{efron1994introduction}, {with supporting results in S3 and S4. The Jackknife estimator $\widehat{\theta}_{JK}$ has similar bias with $\widehat{\theta}_{S}$ but with increased MSE.}

{In Table} \ref{sim_tab_cond}, {we also evaluate the conditional bias and MSE given the third treatment group $i=3$ is being selected. The true response mean of this group is larger than or equal to the other two groups under four simulation scenarios specified above. The marginal bias and MSE are evaluated under some additional settings with $I = 4$ treatment groups in Table} \ref{sim_tab_uncond_four}. {Results and conclusions of these two additional analyses are consistent with Table} \ref{sim_tab_uncond}.  

{The overall recommendation is that the double Bootstrap estimators $\widehat{\theta}^{(2)}_{PB}$ and $\widehat{\theta}^{(2)}_{NB}$ can be applied to achieve the smallest bias, but with slightly larger MSE as compared with $\widehat{\theta}$. The hybrid parameters $\widehat{\theta}^{(2)}_{PB, S}$ and $\widehat{\theta}^{(2)}_{NB, S}$ achieve a better balance between bias reduction and MSE reduction. }

\renewcommand{\arraystretch}{1.4}
\begin{table}[ht]
	\centering
	\fontsize{8pt}{8pt}\selectfont
	\begin{tabular}{cc|ccc|cccccc}
		\toprule 
		& & \multicolumn{3}{|c|}{Existing Estimators} & \multicolumn{6}{c}{Proposed Estimators} \\
		Scenario & $\boldsymbol{\theta}$ & $\widehat{\theta}$ &
		$\widehat{\theta}_{S}$ & $\widehat{\theta}^{(1)}_{NB}$ & $\widehat{\theta}^{(2)}_{NB}$ &
		$\widehat{\theta}^{(2)}_{NB, S}$ & $\widehat{\theta}^{(1)}_{PB}$ & $\widehat{\theta}^{(2)}_{PB}$ &
		$\widehat{\theta}^{(2)}_{PB, S}$ & $\widehat{\theta}_{JK}$     \\ 
		\midrule
		S1 & (1, 1, 1) & 0.67 (0.80) & 0.18 (0.35) & 0.41 (0.65) & 0.07 (0.83) & 0.14 ($\underline{0.33}$) & 0.40 (0.65) & $\boldsymbol{0.06}$ (0.83) & 0.14 ($\underline{0.33}$) & 0.35 (1.15) \\ 
		& (1, 1, 1.2) & 0.54 (0.64) & 0.05 (0.32) & 0.27 (0.55) & -0.06 (0.83) & $\boldsymbol{0.01}$ ($\underline{0.31}$) & 0.27 (0.55) & -0.07 (0.84) & $\boldsymbol{0.01}$ ($\underline{0.31}$) & 0.22 (1.07) \\ 
		& (1, 1.1, 1.2) & 0.58 (0.69) & 0.09 (0.32) & 0.32 (0.58) & -$\boldsymbol{0.02}$ (0.83) & 0.05 ($\underline{0.31}$) & 0.31 (0.58) & -0.03 (0.84) & 0.05 ($\underline{0.31}$) & 0.27 (1.09) \\ 
		& (1, 1.2, 1.2) & 0.60 (0.72) & 0.11 (0.33) & 0.33 (0.60) & $\boldsymbol{0.00}$ (0.84) & 0.07 (0.32) & 0.33 (0.60) & -0.01 (0.85) & 0.07 ($\underline{0.31}$) & 0.27 (1.14) \\ 
		\\
		S2 & (1, 1, 1) & 0.55 (0.55) & 0.26 (0.32) & 0.33 (0.44) & $\boldsymbol{0.05}$ (0.57) & 0.18 ($\underline{0.29}$) & 0.32 (0.44) & $\boldsymbol{0.05}$ (0.58) & 0.18 ($\underline{0.29}$) & 0.28 (0.80) \\ 
		& (1, 1, 1.2) & 0.42 (0.45) & 0.13 (0.30) & 0.21 (0.41) & -0.06 (0.62) & $\boldsymbol{0.05}$ ($\underline{0.28}$) & 0.20 (0.41) & -0.07 (0.62) & $\boldsymbol{0.05}$ (0.29) & 0.17 (0.76) \\ 
		& (1, 1.1, 1.2) & 0.46 (0.49) & 0.17 (0.31) & 0.25 (0.43) & -$\boldsymbol{0.02}$ (0.62) & 0.09 ($\underline{0.29}$) & 0.24 (0.43) & -0.03 (0.62) & 0.09 ($\underline{0.29}$) & 0.20 (0.78) \\ 
		& (1, 1.2, 1.2) & 0.49 (0.51) & 0.20 (0.32) & 0.27 (0.44) & $\boldsymbol{0.00}$ (0.62) & 0.12 ($\underline{0.30}$) & 0.27 (0.44) & -0.01 (0.63) & 0.11 ($\underline{0.30}$) & 0.21 (0.83) \\ 
		\midrule
		Scenario & $w$ & $\widehat{\theta}$ &
		$\widehat{\theta}_{S}$ & $\widehat{\theta}^{(1)}_{NB}$ & $\widehat{\theta}^{(2)}_{NB}$ &
		$\widehat{\theta}^{(2)}_{NB, S}$ & $\widehat{\theta}^{(1)}_{PB}$ & $\widehat{\theta}^{(2)}_{PB}$ &
		$\widehat{\theta}^{(2)}_{PB, S}$ & $\widehat{\theta}_{JK}$     \\ 
		\midrule
		S3 & 0.1 & 0.52 (0.56) & 0.07 (0.27) & 0.27 (0.48) & -$\boldsymbol{0.03}$ (0.68) & $\boldsymbol{0.03}$ ($\underline{0.26}$) & 0.27 (0.48) & -0.04 (0.69) & $\boldsymbol{0.03}$ ($\underline{0.26}$) & 0.22 (0.91) \\ 
		& 0.2 & 0.46 (0.45) & 0.05 (0.22) & 0.24 (0.39) & -0.04 (0.57) & $\boldsymbol{0.02}$ ($\underline{0.21}$) & 0.23 (0.39) & -0.05 (0.57) & $\boldsymbol{0.02}$ ($\underline{0.21}$) & 0.19 (0.73) \\ 
		& 0.3 & 0.43 (0.40) & 0.06 (0.20) & 0.22 (0.35) & -0.03 (0.49) & $\boldsymbol{0.02}$ ($\underline{0.19}$) & 0.22 (0.34) & -0.04 (0.49) & $\boldsymbol{0.02}$ ($\underline{0.19}$) & 0.19 (0.62) \\ 
		& 0.5 & 0.39 (0.39) & 0.07 (0.21) & 0.21 (0.33) & -$\boldsymbol{0.02}$ (0.45) & $\boldsymbol{0.02}$ ($\underline{0.20}$) & 0.20 (0.32) & -0.03 (0.44) & $\boldsymbol{0.02}$ ($\underline{0.20}$) & 0.16 (0.51) \\ 
		\\
		S4 & 0.1 & 0.51 (0.54) & 0.06 (0.26) & 0.26 (0.46) & -0.04 (0.68) & $\boldsymbol{0.03}$ ($\underline{0.25}$) & 0.26 (0.46) & -0.05 (0.68) & $\boldsymbol{0.03}$ ($\underline{0.25}$) & 0.21 (0.90) \\ 
		& 0.2 & 0.45 (0.45) & 0.05 (0.22) & 0.23 (0.38) & -0.04 (0.56) & $\boldsymbol{0.01}$ ($\underline{0.21}$) & 0.23 (0.38) & -0.05 (0.57) & $\boldsymbol{0.01}$ ($\underline{0.21}$) & 0.19 (0.74) \\ 
		& 0.3 & 0.41 (0.37) & 0.04 (0.18) & 0.21 (0.32) & -0.05 (0.48) & $\boldsymbol{0.01}$ ($\underline{0.17}$) & 0.20 (0.32) & -0.06 (0.49) & $\boldsymbol{0.01}$ ($\underline{0.17}$) & 0.16 (0.64) \\ 
		& 0.5 & 0.38 (0.32) & 0.03 (0.16) & 0.19 (0.27) & -0.05 (0.41) & $\boldsymbol{0.00}$ ($\underline{0.15}$) & 0.18 (0.27) & -0.05 (0.41) & $\boldsymbol{0.00}$ ($\underline{0.15}$) & 0.15 (0.54) \\ 
		\bottomrule
	\end{tabular}
	\caption{Marginal bias and MSE in parenthesis of three existing estimators and six proposed estimators. Within each row, the bias with the smallest absolute value is in bold, and the smallest MSE is underlined.}
	\label{sim_tab_uncond}
\end{table}

\begin{table}[ht]
	\centering
	\fontsize{8pt}{8pt}\selectfont
	\begin{tabular}{cc|ccc|cccccc}
		\toprule 
		& & \multicolumn{3}{|c|}{Existing Estimators} & \multicolumn{6}{c}{Proposed Estimators} \\
		Scenario & $\boldsymbol{\theta}$ & $\widehat{\theta}$ &
		$\widehat{\theta}_{S}$ & $\widehat{\theta}^{(1)}_{NB}$ & $\widehat{\theta}^{(2)}_{NB}$ &
		$\widehat{\theta}^{(2)}_{NB, S}$ & $\widehat{\theta}^{(1)}_{PB}$ & $\widehat{\theta}^{(2)}_{PB}$ &
		$\widehat{\theta}^{(2)}_{PB, S}$ & $\widehat{\theta}_{JK}$     \\ 
		\midrule
		S1 & (1, 1, 1) & 0.68 (0.81) & 0.19 (0.35) & 0.41 (0.65) & $\boldsymbol{0.07}$ (0.82) & 0.15 ($\underline{0.33}$) & 0.41 (0.65) & $\boldsymbol{0.07}$ (0.84) & 0.15 ($\underline{0.33}$) & 0.36 (1.13) \\ 
		& (1, 1, 1.2) & 0.58 (0.69) & 0.08 (0.34) & 0.33 (0.60) & 0.02 (0.84) & 0.03 ($\underline{0.33}$) & 0.32 (0.60) & $\boldsymbol{0.01}$ (0.84) & 0.03 (0.34) & 0.29 (1.07) \\ 
		& (1, 1.1, 1.2) & 0.62 (0.74) & 0.12 (0.34) & 0.36 (0.62) & 0.05 (0.84) & 0.07 ($\underline{0.33}$) & 0.36 (0.62) & $\boldsymbol{0.04}$ (0.85) & 0.07 ($\underline{0.33}$) & 0.32 (1.10) \\ 
		& (1, 1.2, 1.2) & 0.63 (0.76) & 0.14 (0.34) & 0.37 (0.63) & 0.05 (0.85) & 0.09 ($\underline{0.32}$) & 0.37 (0.63) & $\boldsymbol{0.04}$ (0.87) & 0.09 ($\underline{0.32}$) & 0.31 (1.15) \\ 
		\\
		S2 & (1, 1, 1) & 0.71 (0.81) & 0.42 (0.49) & 0.50 (0.67) & 0.23 (0.77) & 0.34 ($\underline{0.44}$) & 0.49 (0.67) & $\boldsymbol{0.22}$ (0.78) & 0.34 ($\underline{0.44}$) & 0.44 (1.08) \\ 
		& (1, 1, 1.2) & 0.59 (0.66) & 0.30 (0.42) & 0.40 (0.59) & 0.16 (0.74) & 0.22 ($\underline{0.39}$) & 0.39 (0.59) & $\boldsymbol{0.15}$ (0.74) & 0.22 (0.40) & 0.37 (0.93) \\ 
		& (1, 1.1, 1.2) & 0.64 (0.71) & 0.34 (0.44) & 0.43 (0.61) & $\boldsymbol{0.19}$ (0.75) & 0.26 ($\underline{0.40}$) & 0.43 (0.61) & $\boldsymbol{0.19}$ (0.74) & 0.26 (0.41) & 0.42 (0.92) \\ 
		& (1, 1.2, 1.2) & 0.65 (0.72) & 0.35 (0.45) & 0.44 (0.61) & 0.18 (0.75) & 0.27 ($\underline{0.40}$) & 0.43 (0.62) & $\boldsymbol{0.17}$ (0.75) & 0.27 ($\underline{0.40}$) & 0.39 (1.00) \\
		\midrule
		Scenario & $w$ & $\widehat{\theta}$ &
		$\widehat{\theta}_{S}$ & $\widehat{\theta}^{(1)}_{NB}$ & $\widehat{\theta}^{(2)}_{NB}$ &
		$\widehat{\theta}^{(2)}_{NB, S}$ & $\widehat{\theta}^{(1)}_{PB}$ & $\widehat{\theta}^{(2)}_{PB}$ &
		$\widehat{\theta}^{(2)}_{PB, S}$ & $\widehat{\theta}_{JK}$     \\ 
		\midrule
		S3 & 0.1 & 0.55 (0.61) & 0.10 (0.29) & 0.32 (0.52) & 0.03 (0.70) & 0.06 ($\underline{0.28}$) & 0.31 (0.52) & $\boldsymbol{0.02}$ (0.72) & 0.06 ($\underline{0.28}$) & 0.28 (0.92) \\ 
		& 0.2 & 0.48 (0.48) & 0.07 (0.23) & 0.27 (0.41) & 0.02 (0.57) & 0.03 ($\underline{0.22}$) & 0.27 (0.41) & $\boldsymbol{0.01}$ (0.56) & 0.04 (0.23) & 0.25 (0.72) \\ 
		& 0.3 & 0.45 (0.43) & 0.08 (0.22) & 0.26 (0.37) & 0.02 (0.51) & 0.04 ($\underline{0.21}$) & 0.25 (0.37) & $\boldsymbol{0.01}$ (0.50) & 0.04 ($\underline{0.21}$) & 0.23 (0.62) \\ 
		& 0.5 & 0.41 (0.40) & 0.11 (0.24) & 0.24 (0.34) & 0.04 (0.45) & 0.06 ($\underline{0.22}$) & 0.24 (0.34) & $\boldsymbol{0.03}$ (0.44) & 0.06 ($\underline{0.22}$) & 0.21 (0.51) \\ 
		\\
		S4 & 0.1 & 0.55 (0.59) & 0.10 (0.27) & 0.31 (0.50) & 0.03 (0.69) & 0.06 ($\underline{0.26}$) & 0.31 (0.50) & $\boldsymbol{0.02}$ (0.69) & 0.06 ($\underline{0.26}$) & 0.28 (0.92) \\ 
		& 0.2 & 0.49 (0.49) & 0.08 ($\underline{0.23}$) & 0.28 (0.41) & 0.03 (0.57) & 0.04 ($\underline{0.23}$) & 0.28 (0.42) & $\boldsymbol{0.02}$ (0.57) & 0.04 ($\underline{0.23}$) & 0.25 (0.74) \\ 
		& 0.3 & 0.44 (0.40) & 0.06 (0.19) & 0.25 (0.35) & 0.01 (0.48) & 0.02 ($\underline{0.18}$) & 0.25 (0.34) & $\boldsymbol{0.00}$ (0.49) & 0.02 ($\underline{0.18}$) & 0.22 (0.63) \\ 
		& 0.5 & 0.40 (0.34) & 0.04 (0.17) & 0.22 (0.30) & $\boldsymbol{0.00}$ (0.42) & 0.01 ($\underline{0.16}$) & 0.21 (0.30) & -0.01 (0.43) & 0.01 ($\underline{0.16}$) & 0.18 (0.56) \\ 	
		\bottomrule
	\end{tabular}
	\caption{Conditional bias and MSE in parenthesis of three existing estimators and six proposed estimators. Within each row, the bias with the smallest absolute value is in bold, and the smallest MSE is underlined.}
	\label{sim_tab_cond}
\end{table}

\begin{table}[ht]
	\centering
	\fontsize{8pt}{8pt}\selectfont
	\begin{tabular}{c|ccc|cccccc}
		\toprule 
		& \multicolumn{3}{c|}{Existing Estimators} & \multicolumn{6}{c}{Proposed Estimators} \\
		$\boldsymbol{\theta}$ & $\widehat{\theta}$ &
		$\widehat{\theta}_{S}$ & $\widehat{\theta}^{(1)}_{NB}$ & $\widehat{\theta}^{(2)}_{NB}$ &
		$\widehat{\theta}^{(2)}_{NB, S}$ & $\widehat{\theta}^{(1)}_{PB}$ & $\widehat{\theta}^{(2)}_{PB}$ &
		$\widehat{\theta}^{(2)}_{PB, S}$ & $\widehat{\theta}_{JK}$     \\ 
		\midrule
		(1, 1, 1, 1) & 0.67 (0.80) & 0.19 (0.30) & 0.48 (0.69) & 0.07 (0.86) & 0.13 (0.27) & 0.48 (0.69) & $\boldsymbol{0.06}$ (0.87) & 0.13 ($\underline{0.26}$) & 0.41 (1.30) \\ 
		(1, 1, 1, 1.2) & 0.47 (0.57) & 0.05 (0.27) & 0.34 (0.58) & -0.07 (0.87) & -$\boldsymbol{0.02}$ ($\underline{0.25}$) & 0.34 (0.58) & -0.08 (0.88) & -$\boldsymbol{0.02}$ (0.26) & 0.26 (1.22) \\ 
		(1, 1.05, 1.1, 1.2) & 0.53 (0.63) & 0.09 (0.28) & 0.39 (0.61) & -$\boldsymbol{0.02}$ (0.85) & $\boldsymbol{0.02}$ ($\underline{0.26}$) & 0.38 (0.61) & -0.03 (0.86) & $\boldsymbol{0.02}$ ($\underline{0.26}$) & 0.32 (1.23) \\ 
		(1, 1.1, 1.2, 1.2) & 0.57 (0.68) & 0.12 (0.29) & 0.42 (0.64) & 0.01 (0.85) & 0.05 ($\underline{0.26}$) & 0.41 (0.64) & $\boldsymbol{0.00}$ (0.86) & 0.05 ($\underline{0.26}$) & 0.36 (1.24) \\ 
		\bottomrule
	\end{tabular}
	\caption{Marginal bias and MSE in parenthesis of three existing estimators and six proposed estimators when $I=4$. Within each row, the bias with the smallest absolute value is in bold, and the smallest MSE is underlined.}
	\label{sim_tab_uncond_four}
\end{table}

\subsection{The Choice of $B$ and Higher-Order Bootstrap Methods}
\label{sec:sen_boot}

{In this section, we provide some insights and guidance on how to choose the value of $B$ in Bootstrap methods, and the feasibility of higher-order Bootstrap methods. Under 4 different values of $\boldsymbol{\theta}$ in S1, Table} \ref{sim_tab_boot} {evaluates single, double and triple Bootstrap methods for both parametric and nonparametric versions. Due to computation burdens, triple Bootstrap methods $\widehat{\theta}^{(3)}_{PB}$ and $\widehat{\theta}^{(3)}_{NB}$ are only assessed under $B = 80$ and $100$. }

{For single and double Bootstrap methods, there is limited improvement in bias and MSE when increasing $B = 80$ to $1000$ under scenarios we considered. Therefore, we utilize $B = 80$ in this study, and $B = 1000$ for the real data example in the next section. For other problems, one can implement Bootstrap methods with several values of $B$ to find a proper one with a reasonable computation time. When it comes to higher-order Bootstrap, for example triple Bootstrap methods, their bias and MSE can be even worse than the single Bootstrap version. The high MSE of double Bootstrap estimators is carried over to triple Bootstrap estimators by an additional layer of iteration. With even more intensive computation, triple Bootstrap or even higher-order Bootstrap methods are not recommended for the settings considered. }

\begin{table}[ht]
	\centering
	\fontsize{9.5pt}{9.5pt}\selectfont
	\begin{tabular}{cccccccc}
		\toprule
		$\boldsymbol{\theta}$ & B &  $\widehat{\theta}^{(1)}_{PB}$ & $\widehat{\theta}^{(2)}_{PB}$ &
		$\widehat{\theta}^{(3)}_{PB}$ & $\widehat{\theta}^{(1)}_{NB}$ & $\widehat{\theta}^{(2)}_{NB}$ &
		$\widehat{\theta}^{(3)}_{NB}$  \\ 
		\midrule
		(1, 1, 1) &  80 & 0.40 (0.65) & 0.06 (0.83) & -0.37 (1.97) & 0.40 (0.65) & 0.07 (0.83) & -0.36 (1.94) \\ 
		& 100 & 0.39 (0.63) & 0.05 (0.82) & -0.39 (1.95) & 0.40 (0.63) & 0.06 (0.82) & -0.38 (1.92) \\ 
		& 500 & 0.41 (0.64) & 0.07 (0.81) &  & 0.41 (0.65) & 0.08 (0.81) &  \\ 
		& 1000 & 0.38 (0.62) & 0.04 (0.80) &  & 0.39 (0.63) & 0.05 (0.80) &  \\ 
		\\
		(1, 1, 1.2) &  80 & 0.28 (0.57) & -0.06 (0.84) & -0.49 (2.08) & 0.28 (0.57) & -0.05 (0.84) & -0.47 (2.04) \\ 
		& 100 & 0.27 (0.55) & -0.07 (0.83) & -0.51 (2.07) & 0.27 (0.56) & -0.06 (0.82) & -0.49 (2.03) \\ 
		& 500 & 0.28 (0.56) & -0.06 (0.82) &  & 0.28 (0.57) & -0.05 (0.81) &  \\ 
		& 1000 & 0.26 (0.55) & -0.08 (0.82) &  & 0.27 (0.55) & -0.07 (0.81) &  \\ 
		\\
		(1, 1.1, 1.2) &  80 & 0.31 (0.59) & -0.03 (0.84) & -0.46 (2.06) & 0.31 (0.59) & -0.02 (0.84) & -0.45 (2.03) \\ 
		& 100 & 0.30 (0.57) & -0.04 (0.82) & -0.48 (2.03) & 0.30 (0.57) & -0.04 (0.82) & -0.47 (2.00) \\ 
		& 500 & 0.31 (0.58) & -0.03 (0.81) &  & 0.32 (0.58) & -0.02 (0.81) &  \\ 
		& 1000 & 0.29 (0.57) & -0.05 (0.82) &  & 0.30 (0.57) & -0.04 (0.81) &  \\ 
		\\
		(1, 1.2, 1.2) &  80 & 0.34 (0.61) & 0.00 (0.85) & -0.43 (2.04) & 0.35 (0.61) & 0.02 (0.84) & -0.42 (2.01) \\ 
		& 100 & 0.33 (0.59) & 0.00 (0.82) & -0.44 (2.01) & 0.34 (0.60) & 0.00 (0.82) & -0.43 (1.97) \\ 
		& 500 & 0.35 (0.61) & 0.01 (0.82) &  & 0.35 (0.61) & 0.02 (0.81) &  \\ 
		& 1000 & 0.33 (0.59) & -0.01 (0.82) &  & 0.33 (0.59) & 0.00 (0.81) &  \\ 
		\bottomrule
	\end{tabular}
	\caption{Marginal bias and MSE in parenthesis of single, double and triple Bootstrap methods with varying $B$.}
	\label{sim_tab_boot}
\end{table}

\section{Real Data Example}
\label{sec:real}

AWARD-5 was an adaptive, dose-finding, seamless Phase 2/3 study of dulaglutide for the
treatment of type 2 diabetes mellitus \citep{geiger2012adaptive}. The study had a dose-finding portion (Stage 1) with Bayesian response adaptive randomization to evaluate 7 dulaglutide doses and a fixed scheme (Stage 2) to confirm findings of 2 selected doses (0.75 mg and 1.5 mg) \citep{skrivanek2014dose}. The adaptive randomization at Stage 1 and dose selection at the end of Stage 2 was informed by a clinical utility index (CUI), a single metric that reflects four prespecified safety and efficacy response measures \citep{geiger2012adaptive}. Sample size re-estimation was also performed for Stage 2 based on the data from Stage 1 \citep{geiger2012adaptive, skrivanek2014dose, ctgov}.

For illustration purposes, we consider a simplified problem of treating Stage 1 as a previous Phase 2 study, while Stage 2 as the new Phase 3 study. Our goal is to accurately estimate the response mean of the selected group dulaglutide 1.5 mg to plan its sample size for Stage 2 based on results in Stage 1. The dosing regimen dulaglutide 1.5 mg was selected as the most efficacious group for further testing in Stage 2 during the actual trial conduct of AWARD-5 \citep{skrivanek2014dose}. Assessments are based on the primary efficacy endpoint of change from Baseline (CHG) of glycosylated hemoglobin (HbA1c) at Week 52. For notation consistency, we use the negative of CHG (decrease in HbA1c) with a larger value denoting a better response. Table \ref{T:real_assump} summarizes the response mean (based on Bayesian posterior mean), the sample size, and the standard deviation (based on Normal approximation of Bayesian $95\%$ credible intervals) for each of the 7 active treatment groups in Stage 1 of dose selection \citep{skrivanek2014dose}. Since publicly available results are only summary statistics by group as in Table \ref{T:real_assump}, we apply the single PB $\widehat{\theta}^{(1)}_{PB}$, the double PB $\widehat{\theta}^{(2)}_{PB}$, and the hybrid estimator $\widehat{\theta}^{(2)}_{PB, S}$ to estimate the response mean of 1.5 mg for sample size re-assessment in Stage 2. The Bootstrap sample size is $B = 1,000$, and therefore, $\widehat{\theta}^{(1)}_{PB}$, $\widehat{\theta}^{(2)}_{PB}$ and $\widehat{\theta}^{(2)}_{PB, S}$ require $10^3$, $10^6$ and $10^6$ resamples, respectively. 

{The traditional estimator $\widehat{\theta}$ in} (\ref{equ:theta_naive}) {is $1.33$ as the maximum of -CHGs from 7 active treatment groups. Our three PB estimators are $\widehat{\theta}^{(1)}_{PB} = 1.28$, $\widehat{\theta}^{(2)}_{PB} = 1.20$ and $\widehat{\theta}^{(2)}_{PB, S} = 1.16$, with computational time on a standard laptop of $0.04$ second, $23.1$ seconds and $23.1$ seconds, respectively. These results are consistent with simulation results / conclusions in Section} \ref{sec:sim}, {where $\widehat{\theta}$ has a large positive estimation bias, and $\widehat{\theta}^{(1)}_{PB}$ has a moderate positive bias, while $\widehat{\theta}^{(2)}_{PB}$ and $\widehat{\theta}^{(2)}_{PB, S}$ have biases close to zero.}

\begin{table}[ht]
	\centering
	\begin{tabular}{lccc}
		Group & -CHG of HbA1c at Week 52 & n & SD \\ 
		\toprule
		Dulaglutide 0.25 mg & 0.82 & 13 & 0.55 \\
		Dulaglutide 0.5	mg & 0.95 & 16 & 0.42 \\
		Dulaglutide 0.75 mg & 0.93 & 20 & 0.59 \\
		Dulaglutide 1 mg & 1.00 & 8 & 0.40 \\
		Dulaglutide 1.5 mg & 1.33 & 18 & 0.67 \\
		Dulaglutide 2 mg & 1.28 & 24 & 0.49 \\
		Dulaglutide 3 mg & 1.00 & 10 & 0.42 \\
		\hline
		Estimator & Value &  & \\ 
		\hline
		$\widehat{\theta}$ & $1.33$ & & \\
		$\widehat{\theta}^{(1)}_{PB} $ & $1.28$ & & \\ $\widehat{\theta}^{(2)}_{PB} $ & $1.20$ & & \\ $\widehat{\theta}^{(2)}_{PB, S} $ & $1.16$ & &  \\
		\bottomrule
	\end{tabular}
	\caption{Summary statistics of Stage 1 are based on Bayesian posterior means and $95\%$ credible intervals of CHG of HbA1c at Week 52 \citep{skrivanek2014dose}. The standard deviations (SD) are computed by Normal approximation of the $95\%$ credible intervals. {Values of 4 estimators are presented.} }
	\label{T:real_assump}
\end{table}


\section{Discussion}
\label{sec:dis}

We summarize several attributes of those computational methods in Table \ref{T:summary}. PB methods can either use patient-level data or summary results by treatment group, while NB methods and the Jackknife need patient-level data. With the least computational resource, the Jackknife method can give exact results. As compared with PB, NB does not necessarily require specific distribution assumptions. Based on simulation studies with outliers in Section \ref{sec:sim}, PB with the Normal assumption has a satisfactory performance when inferring response means of continuous endpoints. {Bootstrap methods are capable of conducting a second-order resampling to decrease bias. However, double Bootstrap and Jackknife are observed to have larger MSEs than the traditional estimator} $\widehat{\theta}$ {based on results in Section} \ref{sec:sim}. {Generally speaking, correcting the bias may cause a larger increase in variance, and results in a larger MSE} \citep{efron1994introduction}. {The Jackknife method makes a linear approximation to the Bootstrap method, and can be inefficient for nonlinear functions} \citep{efron1994introduction}. {These arguments may explain why both Bootstrap and Jackknife methods can correct bias but with larger MSE than} $\widehat{\theta}$, {and why Jackknife method usually has the largest MSE. Hybrid estimators based on double Bootstrap and shrinkage can reduce both bias and MSE. Therefore, our overall recommendation is to implement PB methods if only summary results are available, and to choose NB methods with patient-level data. Hybrid estimators based on double Bootstrap and shrinkage are preferred to balance reductions in both bias and MSE. }

{The last row of Table} \ref{T:summary} {summarized limitations of our computational methods for bias correction. Single Bootstrap methods have moderate or limited bias reduction, while double Bootstrap methods have increased MSE and require intensive computation. Hybrid estimators also require double Bootstrap with heavy computation. The Jackknife method has limited bias reduction but with increased MSE.}

This article is not intended to completely fill the evidence gap between previous studies and Phase 3 studies. Under a basic scenario where efficacy profiles of the selected group(s) are the same between studies, we show that our computational methods can characterize the efficacy more accurately than the common practice. We have a broader scope with multiple groups for selection and flexibility to accommodate summary results as compared with some previous works. On top of this framework, additional layers can be added to accommodate more complicated problems, e.g., temporal drift and patient heterogeneity. Our framework can be integrated into MCP-Mod \citep{bretz2005combining} to accommodate the dose-ranging part of previous studies. The proposed method can also be broadly applied to other settings, e.g., response-adaptive randomization design with multiple active treatment groups, patient enrichment designs, and other general selection problems. 

In this article, we consider continuous endpoints for illustration. Generalization can be made for binary endpoints, and time-to-event endpoints. Some other future works include targeting treatment differences by adjusting the placebo effect, regression models to accommodate covariates, and improved computational methods to reduce the burden of higher-order Bootstrap methods. 

\begin{table}[ht]
	\centering
	\fontsize{9pt}{9pt}\selectfont
	\renewcommand{\arraystretch}{1.5}
	\begin{tabular}{ p{0.12\linewidth}|p{0.08\linewidth}|p{0.09\linewidth}|p{0.09\linewidth}|p{0.08\linewidth}|p{0.09\linewidth}|p{0.09\linewidth}|p{0.15\linewidth} }
		\multirow{2}{*}{}	 & \multicolumn{3}{c| }{Parametric Bootstrap} & \multicolumn{3}{ c| }{Nonparametric Bootstrap} & Jackknife \\
		\cline{2-8}
		& Single & Double & Hybrid & Single & Double & Hybrid & \\
		\hline
		Notation	& $\widehat{\theta}^{(1)}_{PB}$ & $\widehat{\theta}^{(2)}_{PB}$ &
		$\widehat{\theta}^{(2)}_{PB, S}$ & $\widehat{\theta}^{(1)}_{NB}$ & $\widehat{\theta}^{(2)}_{NB}$ &
		$\widehat{\theta}^{(2)}_{NB, S}$ & $\widehat{\theta}_{JK}$\\
		\hline
		Data source & \multicolumn{3}{p{0.26\linewidth}| }{Subject level data or summary \newline  statistics by groups} & \multicolumn{4}{p{0.41\linewidth}}{Subject level data} \\
		\hline
		Number of \newline sampling \newline iterations & $B$ & $B^2$ & $B^2$ & $B$ & $B^2$ & $B^2$ & $n$ \\
		\hline
		Features & \multicolumn{3}{p{0.26\linewidth}| }{Can utilize summary statistics; \newline Double Bootstrap or hybrid with \newline shrinkage to increase precision} & \multicolumn{3}{p{0.26\linewidth}|}{Free of distribution assumptions; \newline Double Bootstrap or hybrid with \newline shrinkage to increase precision} & \multicolumn{1}{p{0.17\linewidth}}{Exact results; \newline Less computationally \newline intensive} \\
		\hline
		Limitations & Moderate bias \newline reduction & Increased MSE; Intensive computation & Intensive computation & Moderate bias \newline reduction & Increased MSE; Intensive computation & Intensive computation & Moderate bias \newline reduction; \newline Increased MSE \\
		\hline
	\end{tabular}
	\renewcommand{\arraystretch}{1.5}
	\caption{{Summary table of computational methods for bias correction}}
	\label{T:summary}
\end{table}

\section*{Acknowledgements}
The author thanks the Editor, the Associate Editor and two reviewers for their insightful comments. This manuscript was supported by AbbVie Inc. AbbVie participated in the review and approval of the content. Tianyu Zhan is employed by AbbVie Inc., and may own AbbVie stock.

\section*{Supplementary materials}

The R code to replicate results in Section \ref{sec:sim} and \ref{sec:real} is available on GitHub \url{https://github.com/tian-yu-zhan/Bias_Reduction}. Data sharing is not applicable, because simulation is based on results from literature.

\bigskip

\bibliographystyle{Chicago}

\bibliography{TE_ref}

\begin{thebibliography}{}

\bibitem[\protect\citeauthoryear{Bauer, Koenig, Brannath, and Posch}{Bauer
  et~al.}{2010}]{bauer2010selection}
Bauer, P., F.~Koenig, W.~Brannath, and M.~Posch (2010).
\newblock Selection and bias—two hostile brothers.
\newblock {\em Statistics in Medicine\/}~{\em 29\/}(1), 1--13.

\bibitem[\protect\citeauthoryear{Blumenthal and Cohen}{Blumenthal and
  Cohen}{1968}]{blumenthal1968estimation}
Blumenthal, S. and A.~Cohen (1968).
\newblock Estimation of the larger of two normal means.
\newblock {\em Journal of the American Statistical Association\/}~{\em
  63\/}(323), 861--876.

\bibitem[\protect\citeauthoryear{Bretz, Pinheiro, and Branson}{Bretz
  et~al.}{2005}]{bretz2005combining}
Bretz, F., J.~C. Pinheiro, and M.~Branson (2005).
\newblock Combining multiple comparisons and modeling techniques in
  dose-response studies.
\newblock {\em Biometrics\/}~{\em 61\/}(3), 738--748.

\bibitem[\protect\citeauthoryear{Carreras and Brannath}{Carreras and
  Brannath}{2013}]{carreras2013shrinkage}
Carreras, M. and W.~Brannath (2013).
\newblock Shrinkage estimation in two-stage adaptive designs with midtrial
  treatment selection.
\newblock {\em Statistics in Medicine\/}~{\em 32\/}(10), 1677--1690.

\bibitem[\protect\citeauthoryear{ClinicalTrials.gov}{ClinicalTrials.gov}{2015}]{ctgov}
ClinicalTrials.gov (2015).
\newblock {A Study of LY2189265 Compared to Sitagliptin in Participants With
  Type 2 Diabetes Mellitus on Metformin}.
\newblock . \url{https://clinicaltrials.gov/ct2/show/NCT00734474}.

\bibitem[\protect\citeauthoryear{Dahiya}{Dahiya}{1974}]{dahiya1974estimation}
Dahiya, R.~C. (1974).
\newblock Estimation of the mean of the selected population.
\newblock {\em Journal of the American Statistical Association\/}~{\em
  69\/}(345), 226--230.

\bibitem[\protect\citeauthoryear{Davison and Hinkley}{Davison and
  Hinkley}{1997}]{davison1997bootstrap}
Davison, A.~C. and D.~V. Hinkley (1997).
\newblock {\em Bootstrap methods and their application}.
\newblock Cambridge University Press.

\bibitem[\protect\citeauthoryear{Efron and Tibshirani}{Efron and
  Tibshirani}{1994}]{efron1994introduction}
Efron, B. and R.~J. Tibshirani (1994).
\newblock {\em An introduction to the bootstrap}.
\newblock CRC Press.

\bibitem[\protect\citeauthoryear{{Food and Drug Administration}}{{Food and Drug
  Administration}}{2017}]{fda2017}
{Food and Drug Administration} (2017).
\newblock {22 Case Studies Where Phase 2 and Phase 3 Trials Had Divergent
  Results}.
\newblock \url{https://www.fda.gov/media/102332/download}.

\bibitem[\protect\citeauthoryear{Geiger, Skrivanek, Gaydos, Chien, Berry,
  Berry, and Anderson~Jr}{Geiger et~al.}{2012}]{geiger2012adaptive}
Geiger, M.~J., Z.~Skrivanek, B.~Gaydos, J.~Chien, S.~Berry, D.~Berry, and J.~H.
  Anderson~Jr (2012).
\newblock An adaptive, dose-finding, seamless phase 2/3 study of a long-acting
  glucagon-like peptide-1 analog (dulaglutide): trial design and baseline
  characteristics.
\newblock {\em Journal of Diabetes Science and Technology\/}~{\em 6\/}(6),
  1319--1327.

\bibitem[\protect\citeauthoryear{Hsieh}{Hsieh}{1981}]{hsieh1981estimating}
Hsieh, H.-K. (1981).
\newblock On estimating the mean of the selected population with unknown
  variance.
\newblock {\em Communications in Statistics-Theory and Methods\/}~{\em
  10\/}(18), 1869--1878.

\bibitem[\protect\citeauthoryear{Hsu, Lau, Fung, and Ulveling}{Hsu
  et~al.}{1986}]{hsu1986monte}
Hsu, Y.-S., K.-N. Lau, H.-G. Fung, and E.~F. Ulveling (1986).
\newblock Monte carlo studies on the effectiveness of the bootstrap bias
  reduction method on 2sls estimates.
\newblock {\em Economics Letters\/}~{\em 20\/}(3), 233--239.

\bibitem[\protect\citeauthoryear{Hwang}{Hwang}{1993}]{hwang1993empirical}
Hwang, J.~T. (1993).
\newblock Empirical bayes estimation for the means of the selected populations.
\newblock {\em Sankhy{\=a}: The Indian Journal of Statistics, Series A\/},
  285--304.

\bibitem[\protect\citeauthoryear{{ICH Guideline E8}}{{ICH Guideline
  E8}}{2022}]{e8}
{ICH Guideline E8} (2022).
\newblock {ICH guideline E8 (R1) on general considerations for clinical
  studies}.
\newblock
  \url{https://www.ema.europa.eu/en/documents/scientific-guideline/ich-e-8-general-considerations-clinical-trials-step-5_en.pdf}.

\bibitem[\protect\citeauthoryear{Ishwaei~D, Shabma, and
  Krishnamoorthy}{Ishwaei~D et~al.}{1985}]{ishwaei1985non}
Ishwaei~D, B., D.~Shabma, and K.~Krishnamoorthy (1985).
\newblock Non-existence of unbiased estimators of ordered parameters.
\newblock {\em Statistics: A Journal of Theoretical and Applied
  Statistics\/}~{\em 16\/}(1), 89--95.

\bibitem[\protect\citeauthoryear{Kosmidis}{Kosmidis}{2014}]{kosmidis2014bias}
Kosmidis, I. (2014).
\newblock Bias in parametric estimation: reduction and useful side-effects.
\newblock {\em Wiley Interdisciplinary Reviews: Computational
  Statistics\/}~{\em 6\/}(3), 185--196.

\bibitem[\protect\citeauthoryear{Kumar and Sharma}{Kumar and
  Sharma}{1993}]{kumar1993unbiased}
Kumar, S. and D.~Sharma (1993).
\newblock Unbiased inestimability of the larger of two parameters.
\newblock {\em Statistics: A Journal of Theoretical and Applied
  Statistics\/}~{\em 24\/}(2), 137--142.

\bibitem[\protect\citeauthoryear{Liang, Wu, Mo, Zhou, Shen, Wang, and
  Zheng}{Liang et~al.}{2019}]{liang2019comparison}
Liang, F., Z.~Wu, M.~Mo, C.~Zhou, J.~Shen, Z.~Wang, and Y.~Zheng (2019).
\newblock {Comparison of treatment effect from randomised controlled phase II
  trials and subsequent phase III trials using identical regimens in the same
  treatment setting}.
\newblock {\em European Journal of Cancer\/}~{\em 121}, 19--28.

\bibitem[\protect\citeauthoryear{Lindley}{Lindley}{1962}]{Lindley1962}
Lindley, D.-V. (1962).
\newblock {Discussion of Professor Stein’s paper "Confidence sets for the
  mean of a multivariate normal distribution"}.
\newblock {\em Journal of the Royal Statistical Society, Series B\/}~{\em 24},
  285–287.

\bibitem[\protect\citeauthoryear{MacKinnon and Smith~Jr}{MacKinnon and
  Smith~Jr}{1998}]{mackinnon1998approximate}
MacKinnon, J.~G. and A.~A. Smith~Jr (1998).
\newblock Approximate bias correction in econometrics.
\newblock {\em Journal of Econometrics\/}~{\em 85\/}(2), 205--230.

\bibitem[\protect\citeauthoryear{Miller}{Miller}{1974}]{miller1974jackknife}
Miller, R.~G. (1974).
\newblock The jackknife-a review.
\newblock {\em Biometrika\/}~{\em 61\/}(1), 1--15.

\bibitem[\protect\citeauthoryear{Ouysse}{Ouysse}{2011}]{ouysse2011computationally}
Ouysse, R. (2011).
\newblock Computationally efficient approximation for the double bootstrap mean
  bias correction.
\newblock {\em Economics Bulletin\/}~{\em 31\/}(3), 2388--2403.

\bibitem[\protect\citeauthoryear{Quenouille}{Quenouille}{1949}]{quenouille1949approximate}
Quenouille, M.~H. (1949).
\newblock Approximate tests of correlation in time-series 3.
\newblock {\em Mathematical Proceedings of the Cambridge Philosophical
  Society\/}~{\em 45}, 483--484.

\bibitem[\protect\citeauthoryear{Robertson, Choodari-Oskooei, Dimairo, Flight,
  Pallmann, and Jaki}{Robertson et~al.}{2023a}]{robertson2023point1}
Robertson, D.~S., B.~Choodari-Oskooei, M.~Dimairo, L.~Flight, P.~Pallmann, and
  T.~Jaki (2023a).
\newblock Point estimation for adaptive trial designs i: A methodological
  review.
\newblock {\em Statistics in Medicine\/}~{\em 42\/}(2), 122--145.

\bibitem[\protect\citeauthoryear{Robertson, Choodari-Oskooei, Dimairo, Flight,
  Pallmann, and Jaki}{Robertson et~al.}{2023b}]{robertson2023point2}
Robertson, D.~S., B.~Choodari-Oskooei, M.~Dimairo, L.~Flight, P.~Pallmann, and
  T.~Jaki (2023b).
\newblock Point estimation for adaptive trial designs ii: practical
  considerations and guidance.
\newblock {\em Statistics in Medicine\/}.

\bibitem[\protect\citeauthoryear{Rosenkranz}{Rosenkranz}{2014}]{rosenkranz2014bootstrap}
Rosenkranz, G.~K. (2014).
\newblock Bootstrap corrections of treatment effect estimates following
  selection.
\newblock {\em Computational Statistics \& Data Analysis\/}~{\em 69}, 220--227.

\bibitem[\protect\citeauthoryear{Saville, Berry, Berry, Viele, and
  Berry}{Saville et~al.}{2022}]{saville2022bayesian}
Saville, B.~R., D.~A. Berry, N.~S. Berry, K.~Viele, and S.~M. Berry (2022).
\newblock {The Bayesian time machine: accounting for temporal drift in
  multi-arm platform trials}.
\newblock {\em Clinical Trials\/}~{\em 19\/}(5), 490--501.

\bibitem[\protect\citeauthoryear{Skrivanek, Gaydos, Chien, Geiger, Heathman,
  Berry, Anderson, Forst, Milicevic, and Berry}{Skrivanek
  et~al.}{2014}]{skrivanek2014dose}
Skrivanek, Z., B.~Gaydos, J.~Chien, M.~Geiger, M.~Heathman, S.~Berry,
  J.~Anderson, T.~Forst, Z.~Milicevic, and D.~Berry (2014).
\newblock Dose-finding results in an adaptive, seamless, randomized trial of
  once-weekly dulaglutide combined with metformin in type 2 diabetes patients
  (award-5).
\newblock {\em Diabetes, Obesity and Metabolism\/}~{\em 16\/}(8), 748--756.

\bibitem[\protect\citeauthoryear{Stallard and Todd}{Stallard and
  Todd}{2005}]{stallard2005point}
Stallard, N. and S.~Todd (2005).
\newblock Point estimates and confidence regions for sequential trials
  involving selection.
\newblock {\em Journal of Statistical Planning and Inference\/}~{\em 135\/}(2),
  402--419.

\bibitem[\protect\citeauthoryear{Vellaisamy and Sharma}{Vellaisamy and
  Sharma}{1988}]{vellaisamy1988estimation}
Vellaisamy, P. and D.~Sharma (1988).
\newblock Estimation of the mean of the selected gamma population.
\newblock {\em Communications in Statistics-Theory and Methods\/}~{\em
  17\/}(8), 2797--2817.

\bibitem[\protect\citeauthoryear{Whitehead}{Whitehead}{1986}]{whitehead1986bias}
Whitehead, J. (1986).
\newblock On the bias of maximum likelihood estimation following a sequential
  test.
\newblock {\em Biometrika\/}~{\em 73\/}(3), 573--581.

\bibitem[\protect\citeauthoryear{Zia, Siu, Pond, and Chen}{Zia
  et~al.}{2005}]{zia2005comparison}
Zia, M.~I., L.~L. Siu, G.~R. Pond, and E.~X. Chen (2005).
\newblock {Comparison of outcomes of phase II studies and subsequent randomized
  control studies using identical chemotherapeutic regimens}.
\newblock {\em Journal of Clinical Oncology\/}~{\em 23\/}(28), 6982--6991.

\end{thebibliography}
\end{document}